\documentstyle [12pt] {article}

\def\f{\frac}

\def\Bb{\bar{B}}

\def\a{\alpha}
\def\b{\beta}
\def\d{\delta}
\def\ep{\epsilon}	

\def\f{\phi}	
\def\fb{\bar{\phi}}
\def\tf{\tilde{\phi}}
\def\tfb{\bar{\tilde{\phi}}}

\def\g{\gamma}

\def\k{\kappa}	
\def\m{\mu}
\def\n{\nu}

\def\th{\theta}			
\def\r{\rho}			
\def\t{\tau}
\def\bt{\bar{\tau}}

\def\D{\Delta}

\def\J{\Psi}
\def\L{\Lambda}

\def\cd{{\cal D}}
\def\cdb{\bar{\cal D}}

\def\cf{{\cal F}}

\def\tcf{\tilde{\cal F}}
\def\tcfb{\bar{\tilde{\cal F}}}

\def\ch{{\cal H}}
\def\ci{{\cal I}}

\def\cl{{\cal L}}

\def\cn{{\cal N}}

\def\cp{{\cal P}}
\def\cpb{\bar{\cal P}}

\def\car{{\cal R}}

\def\cw{{\cal W}}
\def\cwb{\bar{\cal W}}
\def\tcw{\tilde{\cal W}}
\def\tcwb{\bar{\tilde{\cal W}}}

\def\da{\dot{\alpha}}
\def\db{\dot{\beta}}
\def\dg{\dot{\gamma}}
\def\pa{\partial}
\def\half{\frac{1}{2}}
\def\tr{\hbox{Tr}}			
\def\({\left(}
\def\){\right)}
\def\l[{\left[}
\def\r]{\right]}
\def\la|{\left|}
\def\ra|{\right|}
\def\le{\left.}
\def\ri{\right.}

\def\rt{\tilde{r}}

\def\dbua{\bar{D}^{\da}}
\def\dbla{\bar{D}_{\da}}
\def\dua{D^{\alpha}}
\def\dla{D_{\alpha}}

\def\dbdf{\bar{D} D \f}
\def\ddbfb{D \bar{D} \fb}

\def\bd4{{\bar{D}}^4}
\def\meas{\int d^4x \, d^8 \theta}
\def\cmeas{\int d^4x \, D^4 \bar{D}^4}
\def\icmeas{\int d^4x \, D^4 }

\newcommand{\be}{\begin{equation}}
\newcommand{\ee}{\end{equation}}
\newcommand{\beqs}{\begin{eqnarray}}
\newcommand{\eeqs}{\end{eqnarray}}

\def\ni{\noindent}
\def\nn{\nonumber}

\newcommand{\NPB}[1]{Nucl.\ Phys.\ {\bf B#1}}
\newcommand{\PLB}[1]{Phys.\ Lett.\ {\bf B#1}}

\newcommand{\PRD}[1]{Phys.\ Rev.\ {\bf D#1}}

\newcommand{\bW}{\bar{W}}

\newcommand{\shalf}{\hbox{$\frac{1}{2}$}} 
\newcommand{\squart}{\hbox{$\frac{1}{4}$}} 
\newcommand{\ret}{\nonumber\\}      

\begin{document}

\begin{titlepage}

\begin{flushright}
\begin{tabular}{l} ITP-SB-97-53 \\ hep-th/9810152 \\ October, 1997
\end{tabular}
\end{flushright}

\vspace{8mm}
\begin{center} 
{\Large \bf Self-Dual Effective Action of $N$=4 Super-Yang Mills}

\vspace{20mm}

F. Gonzalez-Rey, \footnote{email: glezrey@insti.physics.sunysb.edu},
B. Kulik, \footnote{email:bkulik@sbhep.physics.sunysb.edu}
I.Y. Park  \footnote{email: ipark@insti.physics.sunysb.edu},         
and M. Ro\v{c}ek \footnote{email: rocek@insti.physics.sunysb.edu}

\vspace{4mm} Institute for Theoretical Physics \\ 
State University of New York	\\ 
Stony Brook, N. Y. 11794-3840 \\

\vspace{20mm}

\begin{abstract}

The full low energy effective action of $N=4$ SYM is believed to be
self-dual. Starting with the first two leading terms in a momentum
expansion of this effective action, we perform a duality
transformation and find the conditions for self-duality. These
determine some of the higher order terms. We compare the effective
action of $N=4$ SYM with the probe-source description of type $II_B$
D3-branes in the $AdS_5 \times S_5$ background. We find agreement up to 
six derivative terms if we identify the separation of the 3-branes 
with a redefinition of the
gauge scalar that involves the gauge field strength.

\end{abstract}

\end{center}

\vspace{35mm}

\end{titlepage}
\newpage
\setcounter{page}{1}
\pagestyle{plain}
\pagenumbering{arabic}
\renewcommand{\thefootnote}{\arabic{footnote}} \setcounter{footnote}{0}
\indent

\section{Introduction}

In this paper, we study the consequences of self-duality of $N=4$
Super Yang-Mills (SYM) theory in $N=2$ superspace. The bosonic effective
action of this theory can be compared to the Dirac-Born-Infeld (DBI)
action of gauge theories on branes. The DBI action is self-dual under a
duality that does {\em  not} act on the separation of the branes.
However, the Higgs fields that parameterize this separation are in the
same $N=2$ supermultiplet as the gauge fields, and hence the Higgs
fields that realize $N=2$ supersymmetry {\em linearly} must be related
by a nonlinear gauge-field dependent redefinition to the separation.
This is the most striking consequence of our analysis.

\bigskip

 Duality is a powerful tool for probing strong coupling physics: it
allows us to describe strongly coupled systems using the weakly
coupled Lagrangian of the dual degrees of freedom. The dual 
descriptions of a generic theory may in general be very different. 
There is a however a very special theory, namely $N=4$ Super 
Yang-Mills, which is believed to have isomorphic dual descriptions: 
the electric description has a well defined perturbative expansion
when the gauge coupling $g^2 / 4 \pi$ is weak, while the magnetic 
description is well defined perturbatively when a dual gauge coupling 
$g^2_D / 4 \pi = 4 \pi / g^2$ is weak, {\em i.e.}, the original
coupling is strong. These descriptions are isomorphic in the sense
that the electric effective action written in terms of the electric
field strength and coupling has the same form as the magnetic
effective action written in terms of the magnetic field strength and 
coupling. The theory is also believed to be exactly self-dual when the
gauge coupling takes the value $g^2 = 4 \pi$; then duality leaves the
effective action unchanged. For sake of brevity, in subsequent sections
we use the term self-dual in the broader sense for arbitrary $g^2$. 

 The isomorphy of both descriptions is well understood in the classical 
pure gauge action, where a first order formalism implements the change
from fundamental to dual variables as a Legendre transform. Long ago
it was conjectured that the isomorphy is actually a property of the full 
quantum theory \cite{MO}: the spectrum of BPS states remains the same,
and the quantum effective action of the massless gauge sector has the
same form in electric and magnetic variables. The first
part of this conjecture has been tested after extending the strong-weak 
coupling duality to $SL(2,Z)$ \cite{S}. 

We want to study the consequences of this conjecture by
implementing duality on the $N=4$ SYM effective action. This theory is
a particular case of $N=2$ SYM coupled to adjoint matter, and
formulating it in $N=2$ superspace is useful because the $N=2$
superspace effective action of the massless gauge sector has a well
defined expansion in the external momentum. The leading term in a
momentum expansion is a $N=2$ superpotential (a prepotential in the
terminology of \cite{sw})

\be
 S^{(2)}_{eff} = \ci m \int d^4 x \, D^4 \: \cf (\t, W) 
\ee  

\ni
where $D^4 = D^2 Q^2$ is the chiral measure of $N=2$
superspace\footnote{ $D_{1 \a} = D_\a$, $D_{2 \a} = Q_\a$ are the
supercovariant derivatives associated with the Grassmann coordinates
of $N=2$ superspace. We follow the conventions of \cite{book}.}, $W$
is the $N=2$ gauge field strength and $\t = \th / 2 \pi + i 4 \pi /
g^2$ is the holomorphic gauge coupling. In components $S^{(2)}_{eff}$
gives terms with at most two space-time derivatives. This $N=2$
effective superpotential contains all the divergences and all the
scale and $U(1)_R$ anomalies of the theory. For $N=4$ SYM the
perturbative \cite{ni} and nonperturbative \cite{dorey} quantum
corrections to the tree level superpotential $\cf = \t W^2 / 16 \pi$ 
vanish.

 The next term in the momentum expansion of the $N=2$ effective gauge
action is a $N=2$ nonholomorphic potential integrated with the full 
$N=2$ superspace measure. It is therefore a finite, dimensionless 
real function of the $N=2$ gauge field strengths $W$ and $\bar{W}$. 
Since $\cf$ saturates all the perturbative scale and $U(1)_R$ 
anomalies, the perturbative $N=2$ nonholomorphic potential must be 
scale and $U(1)_R$ invariant. In the massless gauge sector this 
restriction completely fixes the 1-loop $N=2$ nonholomorphic 
potential \cite{ni}

\be
  S^{(4)}_{eff} = \int d^4 x \, d^8 \th \: \ch = \int d^4 x \, d^8 \th \: 
 {c \over 8 \pi} \l[ \ln {W \over \L} + g^0(W) \r]
              \l[ \ln {\bW \over \L} + \bar{g}^0 (\bar{W}) \r] ,
\label{nonhol_kahler}
\ee

\ni
where $g^0$ depends on gauge invariant, scale independent combinations
of the $N=2$ abelian field strengths (for a spontaneously broken 
$SU(N_c)$ gauge theory where we keep only one unbroken $U(1)$ 
background gauge multiplet, we have
$g^0 (W) = 0$). In components $S^{(4)}_{eff}$ contains at most four
space-time derivatives. For scale invariant theories such as $N=4$ SYM,
the abelian nonholomorphic potential is believed to be generated only
at 1-loop \cite{DS}, since higher loop and nonperturbative 
contributions would break the scale invariance of 
$S^{(4)}$. An explicit 1-loop calculation gives the value of the
coefficient $c$ for the abelian piece of $SU(2)$ broken to $U(1)$:
$c = 1 / 2 \pi$ \cite{rikard_vipul, GR, kuz} it was again reproduced 
by comparing $N=1$ components). An 
extension of the analysis in \cite{GR} to the case $SU(N_c)$ broken to 
$SU(N_c-1) \times U(1)$ is included in appendix A. 
Nonperturbative contributions have been studied in \cite{dorey} and 
they indeed give vanishing results.

 Higher derivative contributions $S^{(2n>4)}$ present in the effective 
action of $N=4$ SYM must be also scale and $U(1)_R$ invariant, 
dimensionless and finite.

\bigskip

 We implement the strong-weak coupling duality on this expansion of 
the quantum effective action by using a first order formalism in 
$N=2$ superspace \cite{henning}. In this formulation we relax the 
Bianchi identity constraint 

\be
 D^2_{ab} W = C_{ac} C_{bd} \bar{D}^{2dc} \bW
\label{Bianchi}
\ee

\ni 
and we add a field strength $W_D$ as a Lagrange multiplier that 
enforces the Bianchi identity\footnote{This is very clear when write 
the $N=2$ field strength as the most general superfield obeying the 
Bianchi identity $W_D= \bar{D}^4 D^2_{ab} V_D^{ab}$ \cite{menzi}.} 
on $W$. Then we replace the relaxed chiral superfield $W = \bar{D}^4 V$ 
by its field equation 

\beqs
 S_{eff} & = & {\ci}m \( \icmeas \shalf \, \t W^2 - W W_D\) 
     + \cmeas \cl^{quant} \ret
 {\pa S_{eff} \over \pa W} = 0 & \Longrightarrow & 
      W_D = \t W + 2 i \bd4 \cl^{quant}_{,W} \ret
 {\pa S_{eff} \over \pa \bW} = 0 & \Longrightarrow & 
      \bW_D = \bt \bW - 2 i D^4 \cl^{quant}_{, \bW} \ .
\label{prog}
\eeqs

\ni
Note that the prescribed procedure guarantees that the field equations 
of the original action  

\beqs
 S_{eff} & = & {\ci}m \( \icmeas \shalf \, \t W^2 \) + \cmeas \cl^{quant} \\
 {\pa S_{eff} \over \pa V^{ab}} = & 0 & = D^2_{ab} (\t W + 2 i
 \bd4 \cl^{quant}_{,W} ) + C_{a c} C_{b d} \bar{D}^{2 d c} ( - \bt \bW 
   + 2 i D^4 \cl^{quant}_{, \bW} )  \nn
\eeqs

\ni
become the Bianchi identities of the dual field strength. This
operation is a straightforward generalization to $N=2$ superspace of
the quantum duality performed in \cite{sw}. The $N=2$ SYM theory
studied in \cite{sw} is however very different from $N=4$ SYM: Since 
$S^{(2)}_{eff}$ receives quantum corrections in the $N=2$ SYM case, the 
discussion of duality in \cite{sw} ignored higher derivative terms.
We note further that in contrast to the $N=4$ case, the spectrum of
BPS states is {\em not} $SL(2,Z)$ invariant.
   
\bigskip

 The structure of this article is the following: in section 2 we review
the duality transformation in the classical action of $N=2$ SYM as a
Legendre transform in the path integral of the free theory
\cite{henning}. To illustrate some of the general features we
encounter in these type of transformations we also study a one 
dimensional system where the general form of a self-dual action can be 
found.

 In section 3 we check that the first two terms $S^{(2)}_{eff} + 
S^{(4)}_{eff}$ in the momentum expansion of the $N=4$ low energy 
effective action indeed have the same form in the electric and
magnetic descriptions. We find that the dualization of those two
leading terms produces additional six and higher space-time derivative 
operators in the dual variables $S^{(2n>4)}_{eff}$, which should also 
be present in the original description if the theory is isomorphic
under duality. 
Including such higher derivative terms in the electric effective
action, we find that the dualization gives the same operators in
magnetic variables if their coefficients obey certain relations.

 The effective action of the $U(1)$ gauge background in a spontaneously 
broken $N=4$ $SU(N_c)$ theory, is supposed to describe the dynamics of 
an extremal probe D3-brane in the background of $N_c-1$ overlapping 
extremal source D3-branes \cite{witten}. This can be alternatively
described by a DBI action plus WZ terms. In section 
4 we compare the bosonic degrees of freedom of both actions,
and we find a disagreement that can be resolved up to six-derivative
terms by redefinitions of the $N=1$ superfields in the $N=2$ gauge
multiplet: this reveals the fact that the gauge scalar of the SYM 
theory (in which $N=2$ SUSY is linearly realized) and the separation 
of the 3-branes in the DBI action are not simply proportional to each 
other. We find that this is true only as a first order approximation: 
the relation also involves a nonlinear function of the YM field strengths. 
       
Finally in section 4 we discuss open problems on this line of
research, such as the perturbative/nonperturbative nature of the 
effective action expansion.

\section{Two Simple Examples of self-dual actions}

 Let us begin by reviewing a simple example of the formalism that 
implements the strong-weak coupling duality transformation. We take
the classical $N=2$ Maxwell action in the electric description, and we
relax the Bianchi identity constraint (\ref{Bianchi}) on the field 
strength while at the same time we introduce a Lagrange multiplier as 
we described in the introduction

\be
 8 \pi \: S = \ci m \int d^4x D^4 \; \( \shalf \, \tau W^2 - W W_D \) \ .
\label{multi}
\ee

\ni
integrating out $W_D$ in the path integral of the free theory imposes 
the Bianchi identity (\ref{Bianchi}) on $W$; alternatively, we can 
integrate out $W$, which in this case is the same as replacing it
by its field equation

\be
 W = {W_D \over \t} 
\label{class_dual}
\ee

\ni 
Substituting (\ref{class_dual}) into (\ref{multi}) gives the dual 
action 

\be
 8 \pi \: S_D = \ci m \int d^4 x \, D^4 \; \shalf {-1 \over \t} W_D^2 \ .
\ee

 The original action is therefore equal to the dual one written
with the magnetic variables $W_D$ and $\t_D = -1 / \t$. The classical
Maxwell action is also invariant under a real shift in $\t$, 
$\t \rightarrow \tau + x$, because the shifted integrand is total 
derivative, $\sim \ep^{\a \b \m \n} F_{\a \b} F_{\m \n}$.

 Possible self-dual functionals are actually more general than is 
commonly realized (see for example \cite{gib_rash}. We find this feature 
when we study the
duality of the quantum action. To illustrate this idea consider 
the following one dimensional example: given a function $F$ and 
its Legendre transform

\be     
 \tilde{F} (y) = \le \l[ F(x) - x \, y \r] \ra|_{ x = F_x^{-1} (y) } \ ,
\label{1dmulti}    
\ee      

\ni 
self-duality implies $\tilde{F}(y) = F (y)$ or equivalently      

\be
 F( F_x) = F(x) - x F_x \Rightarrow - x = F_x(F_x) \ . 
\label{xFF}
\ee        

\ni 
The most general solution of eqn. (\ref{xFF}) is $F_x = g^{-1} (ig(x))$ 
where $g$ is any odd function of $x$. Therefore

\be
 F(x) = \int^x dz \, g^{-1} (ig(z)) \ ,
\label{gral_sol}
\ee          

\noindent 
is the most general self-dual function.  In this example the function
$F(x)$ is the analogue of our quantum effective action. Since this
action is constructed as a series expansion, it is useful to refine
the analogy and study the Taylor series expansion of $F(x)$. To do
this, we first Taylor expand $g$

\be
 g = g^{(1)} \,x + {1 \over 3!} g^{(3)} x^3 +{1 \over 5!} g^{(5)} x^5
     + {1 \over 7!} g^{(7)} x^7 + \cdots
\ee  

\ni 
where $g^{(n)} = \le {d^n \over dx^n} g \ra|_{x=0}$. The inverse of 
$g$ is 

\be
 g^{-1}(x) = {1 \over g^{(1)}} x - {1 \over 3!} {g^{(3)} \over
 (g^{(1)})^4} x^3 - {1 \over 5!} { g^{(5)} g^{(1)} - 10 (g^{(3)})^2 \over 
 (g^{(1)})^7} x^5 + \cdots
\ee

\ni 
From the two equations above, one finds

\beqs
 F_x & = & g^{-1}(ig(x)) = ix + {i \over 3} g^{(3)} \, x^3 + {i \over 6} 
  (g^{(3)})^2 \, x^5 + i \( {5 \over 36} {g^{(3)}}^3 - 
  {1 \over 72} g^{(3)} g^{(5)} + {1 \over 2520} g^{(7)} \) x^7 \nn \\
 & & + i \l[ {5 \over 3} g^{(3)} \(- {1 \over 72} g^{(3)} g^{(5)} +
  {1 \over 2520} g^{(7)} \) + {85 \over 648} (g^{(3)})^4 \r] x^9
    + \cdots
\label{coef_dep}
\eeqs

 Here the coefficient of the fifth order term, $i {1 \over 6}
(g^{(3)})^2$, is expressed in terms of $g^{(3)}$, which, up to a numerical 
factor, is the coefficient of third order term. Similarly new 
parameters $g_5$ and $g_7$ appear in the seventh order coefficient,
but the ninth order coefficient is a function of the parameters 
in the lower order coefficients. Furthermore $g_5$ and $g_7$
appear only in the combination $-{1 \over 72} g^{(3)} g^{(5)} +
{1 \over 2520} g^{(7)}$. In the next section, we will observe a
similar behavior for the effective action of $N=4$ SYM.

\section{Dualization of the $N=4$ SYM effective action}

 We have seen that the effective action of $N=4$ SYM up to four
spacetime derivatives is restricted by $U(1)_R$ and scale invariance
to be of the form 

\be
 8 \pi \( S^{(2)}_{eff} + S^{(4)}_{eff} + ... \) = 
  {\ci}m \( \icmeas \; \shalf \t W^2 \) + c \cmeas \: \ln W \ln \bar{W} 
   + ... \ ;
\label{4act}
\ee
 
\ni
for explicit values of the Higgs field, the coefficient $c$ is
calculated in Appendix A.
We are now ready to study if this effective action is self-dual. To 
dualize the action we 
follow the same steps as in (\ref{prog}): we rewrite the action as

\be
 8 \pi \( S^{(2)}_{eff} + S^{(4)}_{eff} + ...\) = {\ci}m \icmeas 
  \( \shalf \t W^2 - W W_D \) + c \meas \ln W \ln \bar{W} + ...\ ,
\label{4dualact}
\ee   

\ni
and the duality equation is given the field equation of the relaxed 
chiral superfield $W$ 

\beqs
 0 & = & \t W - W_D + 2ic {\bd4 \ln \bW \over W} + \dots \ret 
 0 & = & \bt \bW - \bW_D -2ic {D^4 \ln W \over \bW} +
 \dots \label{dualityeq11} \ .
\eeqs

\ni
This is highly nonlinear and hard to invert to find the solution
$W(W_D)$. Notice however that this solution is of the form $W(W_D) =
{W_D \over \t} + higher \ derivatives$. Since we are only studying 
self-duality up to $S^{(4)}$ for the time
being, we can solve (\ref{dualityeq11}) to first order in $D^4$ and
$\bd4$

\beqs
 W & = & {W_D \over \t} \( 1 - 2ic \t {\bd4 \ln \bW_D \over W_D^2}  
             + \dots  \)  \ret      
 \bW & = & {\bW_D \over \bt} \( 1 + 2ic \bt {D^4 \ln W_D \over \bW_D^2 }
           + \dots  \)   
\label{1sol}
\eeqs

\ni 
We substitute (\ref{1sol}) into (\ref{4dualact}) and we find

\beqs
 8 \pi \: S_{eff,D}  & = & 
  {\ci}m \(\icmeas \; \shalf {-1 \over \t} W_D^2 \)  
 + \: c \meas \, \ln W_D \ln \bW_D \ret
& & + \( i c^2 \meas \, \ln \bW_D {\bd4 \ln \bW_D \over {-1 \over \t} W_D^2}  
    \;\; + \; c.c. \)  \; + \; ...
\label{dual_6act}
\eeqs

 Under the map $W_D \rightarrow W$, $\t_D =- {1 \over \t} \rightarrow
\t$ this action is equivalent to (\ref{4act}) up to four spacetime
derivatives. The dual action contains also terms with six spacetime
derivatives proportional to $1/\t_D$. They depend on a
scale and $U(1)_R$ invariant chiral field that contains the
holomorphically normalized field strength $\t W^2$

\be
 B_D = {\bd4 \ln \bW_D \over \t_D W_D^2} = 
 {1 \over 2} {\bd4 \ln \( \t_D \bW_D^2 \) \over \t_D W_D^2} \ .
\ee

\ni
The six derivative term is scale and $U(1)_R$ invariant as a whole. It 
seems therefore that there must be higher 
order terms in the original description of the effective action, which
should also be consistent with the duality conjecture. Let us test
this possibility: we dualize the following effective action 

\beqs
 8 \pi \( S^{(2)}_{eff} + S^{(4)}_{eff} + S^{(6)}_{eff} + ...\) & = & 
     {\ci}m \( \icmeas \; \shalf \t W^2 - W W_D\)    
 + c \meas \, \ln W \ln \bW       \ret
 & & + \(\k \, i c^2 \meas \: \ln \bW {\bd4 \ln \bW \over \tau W^2}
                 \;\; + \; \; c.c. \;\) + ...
\label{6act} 
\eeqs

\ni 
Repeating the same steps as above and solving the field equations 
of $W$ to second order in $\bd4$ and $D^4$ we find

\beqs
 \lefteqn{\!\!\!\!\!\!\!\!\!\!\!\!\!\!\!\!\!\!\!\!
  8 \pi \( S^{(2)}_{eff,D} + S^{(4)}_{eff,D} + S^{(6)}_{eff,D} 
 + S^{(8)}_{eff,D} +...\) = \;
  {\ci}m \icmeas \; \shalf {-1 \over \t} W_D^2  }\ret
 & & \qquad \qquad \qquad + \; c \meas \; \ln W_D \: \ln \bW_D   \\
 & & \qquad \qquad \qquad + \; c^2 \meas \; \( i(1 - \k) \; \ln \bW_D \; 
   {\bd4 \ln \bW_D \over {-1 \over \t} W_D^2} \;\; + \;\; c.c. \; \) \ret
 & & \qquad \qquad \qquad + \; c^3 \meas \; \( (\shalf - \k) \; \ln \bW_D 
         \( {\bd4 \ln \bW_D \over {-1 \over \t} W_D^2} \)^2  \;\; +
                     \;\; c.c. \; \) \ret
 & & \qquad \qquad \qquad \qquad \qquad \qquad + \: (1 - \k- \bar{\k}) \;
       {\bd4 \ln \bW_D \over {-1 \over \t} W_D^2} \;
       {D^4 \ln W_D \over {-1 \over \bt} \bW_D^2} \; + \; ...\nn    
\eeqs                  

 To sixth order, the theory is self-dual
if $\k = \shalf$. For $\k = \shalf$ the eight derivative terms that we 
find in the dual action do not receive
contributions from the dualization of $S^{(2)}_{eff} + S^{(4)}_{eff} +
S^{(6)}_{eff}$. In addition, since at lowest order the duality equation
is still given by $W = W_D / \t$, the scale and $U(1)_R$ invariant
operator depending on $\t W^2$

\beqs
 S^{(8)}_{eff} [\t W^2] & = & c^3 \meas \( \k^{(8)}_2 \; \ln \bW \: 
   \( {\bd4 \ln \bW \over \t W^2} \)^2 \;\; + \;\; c.c. \; \) \ret
& & \qquad \qquad \qquad + \: \k^{(8)}_1 \; 
      {\bd4 \ln \bW \over \t W^2} \; {D^4 \ln W \over \bt \bW^2}  \ , 
\label{8act}
\eeqs

\ni
is dualized at the lowest order into an isomorphic operator depending 
on $- \t_D W_D^2$. Hence, if we set $\k = \shalf$ in (\ref{6act}), we
can add additional terms of the form (\ref{8act}) and the effective 
action is self-dual up to 
$S^{(8)}_{eff}$ for {\em arbitrary} values of the parameters $\k^{(8)}_1$
and $\k^{(8)}_2$.

 The appearance of $S^{(6)}_{eff}$ and $S^{(8)}_{eff}$ illustrates a
key feature of the $N=4$ SYM effective action. Scale and $U(1)_R$
imply that higher derivative operators in the effective action contain
powers of the ratio $\cd = (1 / \t W^2) \bd4$ and its conjugate\footnote{The
holomorphically normalized chiral field strength $\t W^2$ is the
natural object to divide the chiral operator $\bd4$ because the
dimensionless gauge coupling $\t$ maybe promoted to a chiral
superfield with nonvanishing $U(1)_R$ charge \cite{DS}, but the tree
level Lagrangian $\t W^2$ must still transform oppositely to the
measure $D^4$.}. A term with $2 m + 4$ spacetime derivatives will have
$m$ inverse powers of $\t W^2, \bt \bW^2$. Since the dualization of
any operator $S^{(2m+4)}_{eff} [\t W^2, \bt \bW^2]$ gives at lowest
order an isomorphic expression $S^{(2m+4)}_{eff} [-\t_D W_D^2, -\bt_D
\bW_D^2]$, if $m = 2n$ at lowest order the operator is mapped to
itself under duality, while if $m=2n+1$ there is an additional minus
sign.  $S^{(2m+4)}_{eff,D}$ receives additional contributions that we
denote as $\D^{(2m+4)}$. They come from the dualization of lower order
operators $S^{(l<2m+4)}_{eff}$.

\beqs
 S^{(4n+4)}_{eff,D} & = & S^{(4n+4)}_{eff} [\t_D W^2_D] 
                   + \D^{(4n+4)} [\t_D W^2_D]               \ret
 S^{(4n+6)}_{eff,D} & = & - S^{(4n+6)}_{eff} [\t_D W^2_D] 
                   + \D^{(4n+6)} [\t_D W^2_D]
\label{dual_condit}
\eeqs

\ni
Hence, self-duality requires 
$\D^{(4n+4)} = 0$ and $\D^{(4n+6)} = 2 \times S^{(4n+6)}_{eff}$. 
This means that all the coefficients of even order terms are 
completely arbitrary, while the coefficients of odd order operators 
are constrained to be some linear combination of the coefficients 
appearing in lower order terms. This result is analogous to that found 
in the one-dimensional example (\ref{coef_dep}) we introduced in the 
previous section.

 Although we can continue this procedure to construct a self-dual effective
action, inverting 
the duality equations to obtain $W(W_D)$ can be cumbersome at higher 
orders. An equivalent calculation which gives the same results, but 
makes it unnecessary to invert the duality equations is the following:
instead of solving for $W=W(W_D)$, we directly substitute the duality
equation $W_D=W_D(W)$ in $S_{eff} [\t W^2]$ and impose self-duality

\be
S_{eff,D} [\t_D W_D^2] \equiv S_{eff} [\t W^2] - W W_D \ .
\ee   

 The result of this process is an $N=4$ SYM effective action
consistent with the duality conjecture, which depends on powers of the
scale and $U(1)_R$ invariant chiral operators $B$, $\cd$ and of their 
conjugates. The details of the calculation are included in appendix B.

\bigskip

 We seem to have found a recipe to construct a $N=4$ SYM effective
action consistent with the duality conjecture, order by order in a
momentum expansion which is at the same time an expansion in the gauge
coupling. This is a direct consequence of the fact that the actual
expansion variables are the scale and $U(1)_R$ invariant ratio $\cd$ 
and its complex conjugate.

 However these terms are not the most general scale and $U(1)_R$
invariant variables we might consider. We could use for instance a
``square root'' $\cp^{ab}$ of the operator $\cd$ and a corresponding field
$E^{ab}$:

\be
 \cp^{ab} = {\bar{D}^{2 a b} \over \sqrt{\t} W } \ , \; \; 
 E^{ab} =  {\bar{D}^{2 a b} \ln \bW \over \sqrt{\t} W }
    =  {\bar{D}^{2 a b} \ln (\sqrt{\t} \bW) \over \sqrt{\t} W } 
\label{simp_ops}
\ee

\ni
and their complex conjugates. Other possible variables are complex linear
superfields such as

\be
 \bar{D} \( {D \over \sqrt{\bt} \bW } \ln W \) \ ,
\ee

\ni
or fermion bilinears of the form

\be
 (\bar{D} \ln \bW ) \( {\bar{D} \over \sqrt{\t} W } \ln \bW \) \ .
\ee

\ni
Consequently we may form six and higher spacetime derivative terms 
other than the ones we have seen so far. 

 We want to explore if it is possible to construct an $N=4$ SYM 
effective action that contains the more general 
variables. For simplicity we restrict our analysis to the simple
variables in (\ref{simp_ops}). 

 From our previous construction we know that any six derivative term
depending on $\t W^2$ will be dualized at lowest order to the same six
derivative term up to a sign. The dualization of $S^{(2)}_{eff}$ and
$S^{(4)}_{eff}$ gives contributions only to a six derivative term of
the form (\ref{dual_6act}) and thus $S^{(6)}_{eff} [E^{ab},\cp^{ab}]$
must be dualized to an isomorphic $S^{(6)}_{eff,D} [E^{ab}_D,\cp^{ab}_D]$ 
with positive sign. This last condition restricts the form of 
$S^{(6)}_{eff}$ to be

\beqs
 8 \pi \: S^{(6)}_{eff} ( 1/\t, 1/|\t|)& = & \meas \(i {c^2 \over 2} \; 
    \ln \bW \: {\bd4 \ln \bW \over \tau W^2} \;\; +\;\; c.c. \; \) \ret 
& & \qquad \qquad + \: c^2 {\lambda^{(6)} \over \sqrt{\t \bt}} \;  
 {\bar{D}^{2 a b} \ln \bW \over W } \; {D^2_{a b} \ln W \over \bW } \ .
\label{addit_op}
\eeqs

 Extending our construction to even higher derivative operators we 
find more terms mixing the $B$ and $E^{ab}$ variables as 
we increase the order in the expansion. As before, the coefficients of 
terms which are ``even'' under duality remain free parameters, whereas 
those of ``odd'' terms become linear combinations of the lower order 
coefficients. Details of such calculation are given in the appendix B.

\section{SYM effective action and 3-brane dynamics}

The effective action of the abelian piece in $N=4$ $SU(N_c-1) \times
U(1)$ is believed to describe the dynamics of extremal 3-branes in ten
dimensions \cite{witten}: the $N=0$ massless gauge scalar $\f=W|$ is
related to the separation of one of the D3-branes from the rest in one
of the complexified transversal directions, while the $N=0$ massless
gauge field strength $D_{(\a} Q_{\b)} W| = - D_{(\a} \cw_{\b)}| = -
\cf_{\a \b}$ describes the gauge degrees of freedom\footnote{We follow
the conventions of \cite{book} to define the field strength in SL(2,C)
spinor index notation $\cf_{\a \da, \b \db} = C_{\da \db} \cf_{\a \b}
+ C_{\a \b} \bar{\cf}_{\da \db}$.} living in the D3-brane. In this 
section we want to establish the exact nature of this correspondence 
by comparing the bosonic components of the $N=4$ SYM effective action 
(at fixed $<\f> \neq 0$), with the world volume field theory that 
describes the dynamics of a probe D3-brane at a short distance (small 
$\a' <\f>$ \cite{maldac}) from $N_c-1$ source D3-branes \cite{dkps}. 
The effective action for the fields living in the 3-brane is given by a
supersymmetric DBI \cite{aga} \cite{park_rey} action\footnote{We choose
a background in which the RR scalar and 2-form are zero to simplify
our comparison, and therefore the WZ term does not give a contribution
to the gauge effective action. In addition the vacuum angle is set to 
zero, {\em i.e.}, $\t = i 4 \pi / g^2$.}

\be
 S^{bos}_{BI} = \int d^4 x T_3 \(h^{-1} - \sqrt{ - \det ( g_{ab} 
         + F_{ab} - h^{\half} \pa_a X^i \pa_b X^i ) } \)
\label{Born_In}
\ee

\ni
where $x^a , a=0,..,3$ are the coordinates along the 3-brane, $X^i,
i=4,..9$ are the transversal coordinates, $F_{ab}$ is a gauge field
strength on the brane, $T_3 \sim (\a')^{-2}$ is the 3-brane tension and 
$g_{ab} = h^{-\half} \eta_{ab}$ is the induced flat metric of the probe 
3-brane (the normalization $h = 1 + R^4/r^4 = 1 + {8 \cn \over T_3 
(4 \pi X^i X^i)^2}$ is induced by the extremal ten-dimensional background 
metric\footnote{For $N_c -1$ overlapping 3-branes separated from the
last one, we have $\cn = (N_c - 1)$ \cite{chep}.} \cite{hor_stro}).

 The $N=4$ SYM effective action that we have derived is completely 
determined up to terms with four space-time derivatives: the coefficient 
of $S^{(4)}_{eff}$ is given by an explicit 1-loop calculation $c =
(N_c-1)/ 2 \pi$ (see appendix A). In
addition, self-duality fixes the coefficient of some terms 
with six space time derivatives. We can therefore attempt to compare the
bosonic components of $S^{(2)}_{eff} + S^{(4)}_{eff} + S^{(6)}_{eff}$ 
(setting auxiliary fields equal to zero) with the DBI action in 
(\ref{Born_In}).     

 This comparison has been performed for $S^{(2)}_{eff} + S^{(4)}_{eff}$ 
in a $SU(2)$ theory \cite{rikard_vipul}, setting $F_{\a \b}=0$ and focusing
on the velocity dependent terms. Comparisons for various Dp-branes in 
different backgrounds have appeared in \cite{iuri} \cite{esko} \cite{paban}.  

From reference \cite{rikard_vipul} we learn that we can identify the 
separation of the 3-branes in two of the transversal directions with the 
canonically normalized gauge scalar   

\be
 X_i X_i = {1 \over g^2 T_3} \f \fb \; , \; i=4,5 \;, \; \cn = 2 \pi c \ .
\label{corresp}
\ee

\ni
Then we find an agreement between the velocity terms of the Taylor
expanded BI action 

\be
 S_{BI} = \int d^4 x T_3 h^{-1} \( 1 - \sqrt{ 1 - h \dot{X}^i
 \dot{X}^i } \, \) = {T_3 \over 2} \dot{X}^i \dot{X}^i \, + \, {T_3 h
 \over 8} \( \dot{X}^i \dot{X}^i \)^2 + \dots \ ,
\label{exp_BI}
\ee

\ni
and the SYM effective action\footnote{Here we follow \cite{rikard_vipul}
ignoring the spatial dependence of the gauge scalars.}
 
\be
 \( S^{(2)}_{\f} + S^{(4)}_{\f} \) =  
 \int d^4 x \l[ {1 \over 2 g^2} \dot{\f} \dot{\fb} + 
  {c \over 8 \pi} \( { \la| \dot{\f} \ra|^4 \over \la| \f \ra|^4 } - 
  { \ddot{\f} \dot{\fb}^2 \over \f \fb^2 } - 
 { \ddot{\fb} \dot{\f}^2 \over \fb \f^2 } + 
 2 {\la| \ddot{\f} \ra|^2 \over \la| \f \ra|^2 } \)  \r]  \ .
\label{bos4act}
\ee

\ni
In \cite{rikard_vipul} it was pointed out that because of the
identification (\ref{corresp}) the SYM effective action 
contains acceleration terms $\ddot{\f}$, which are absent in the expansion 
of (\ref{exp_BI}). The explanation suggested is that we should put the
action on shell and compare S-matrix elements as opposed to the
effective actions of both descriptions. 

 We establish the correspondence of the SYM effective action and the
DBI action by redefining the $N=1$ component superfields of
$S^{(2)}_{eff} + S^{(4)}_{eff}$ in such a way that the undesired
acceleration terms of the bosonic action are absorbed in the gauge
scalar and the gauge field strength. We conjecture that the
redefinition can be extended to eliminate the acceleration terms of
$S^{(6)}_{eff}$ and even higher derivative terms. Then 
the effective action of $N=4$ SYM written with our redefined $N=1$
superfields becomes the $N=1$ superspace extension of the 3-brane
action. In this redefined action the second supersymmetry must be
realized nonlinearly, but this is precisely what we expect \cite{aga}.
 
 Let us begin by writing the four derivative terms in $N=1$ superspace. 
After integrating the Grassmann coordinates of the second
supersymmetry, integration by parts and some manipulation gives

\beqs
 \lefteqn{S^{(4)}_{eff} =  {c \over 8 \pi} \int d^4 x D^2 \bar{D}^2 
   \le \( Q^2 \ln W \: \bar{Q}^2 \ln \bW - Q^\a \ln W \: (i \pa_{\a\da}) 
                 \bar{Q}^{\da} \ln \bW + \ln W \Box \ln \bW \) \ra|} \ret 
 & & = {c \over 8 \pi} \int \! d^4 x \, d^4 \th 
  \( \l[ { \bar{D}^2 \fb \over \f} - {\cw^2 \over \f^2} \r] \! \! 
  \l[ { D^2 \f \over \fb} - {\cwb^2 \over \fb^2} \r] 
  - D_\a \! \l[ { \cw^\a \over \f} \r] \! \bar{D}_{\da} \! 
   \l[ { \cwb^{\da} \over \fb} \r] + D^2 \ln \f \, \bar{D}^2 \ln \fb \) \ret
 & & = {c \over 8 \pi} \int d^4 x d^4 \th \(
 \bar{D}^2 \fb \l[ { D^2 \f \over \f \fb} - {\cwb^2 \over \f \fb^2}
  - {(D \f)^2 \over \f^2 \fb} \r] + \l[ { \bar{D}^2 \fb \over \f \fb} - 
    {\cw^2 \over \fb \f^2} - {(\bar{D} \fb)^2 \over \fb^2 \f} \r] D^2 \f 
 \ri \ret \ret
& & \qquad \qquad \qquad \qquad - \dla \cw^{\a} \l[ \squart {\dbla \cwb^{\da} 
      \over \f \fb} - \squart {\dla \cw^\a \over \f \fb}  
        + \shalf{ \cwb^{\da} \dbla \fb \over \f \fb^2}  
        -  \shalf{ \cw^{\a} \dla \f \over \fb \f^2} \r] \ret  
& & \qquad \qquad \qquad \qquad - \dbla \cwb^{\da} \l[ \squart {\dla \cw^\a 
      \over \f \fb} - \squart {\dbla \cwb^{\da} \over \f \fb} 
        + \shalf{ \cw^{\a} \dla \f \over \fb \f^2} 
        - \shalf{ \cwb^{\da} \dbla \fb \over \f \fb^2} \r] \ret
& & \qquad \qquad \qquad \qquad \le + \: { \cw^2 \cwb^2 
     + (D \phi)^2 (\bar{D} \fb)^2 - \cw^{\da} \dla \f \cwb^{\da} \dbla \fb 
        \over \f^2 \fb^2} \; \)\ . 
\eeqs

 This particular way of writing the SYM effective action in $N=1$
superspace is very illuminating: in the last line we have terms with 
four fermionic superfields that we will label $S^{(4)}_{fermi}$.
Their $N=0$ bosonic components contain only powers of the velocity and
gauge field strength, but not their spacetime derivatives. In addition, 
the remainder of $S^{(4)}_{eff}$ can be written as  
$\int d^4 \th \fb \bar{D}^2 \D \bar{\J} + \int d^2 \th \cw^{\a} \dla i
\D V + c.c.$ where $\D V$ is real. The unconstrained $N=1$ 
superfields 

\beqs
 \D \bar{\J} & = &  { D^2 \f \over \f \fb} - {\cwb^2 \over \f \fb^2}
  - {(D \f)^2 \over \f^2 \fb}  \label{prepot}\\ 
 i \D V & = & \squart {\dbla \cwb^{\da} \over \f \fb} - 
    \squart {\dla \cw^\a \over \f \fb} + \shalf{ \cwb^{\da} \dbla \fb
     \over \f \fb^2} - \shalf{ \cw^{\a} \dla \f \over \fb \f^2} \nn \ , 
\eeqs

\ni
that contain the acceleration terms of $S^{(4)}_{eff}$ may therefore
be absorbed in a redefinition of the superfields appearing in the 
kinetic term

\beqs
 S^{(2)}_{eff} + S^{(4)}_{eff} & = & 
   {1 \over 2 g^2} \int d^4 x \, d^4 \th \; \tfb \tf 
 - {2 g^2 c^2 \over (8 \pi)^2} \int d^4 x \, d^4 \th \;
           (D^2 \, \D \J) (\bar{D}^2 \, \D \bar{\J}) \\ 
 & & + {1 \over 4 g^2} \int d^4 x D^2 {\tcw^{\a} \tcw_{\a} \over 2} 
 - {4 g^2 c^2 \over 2 (8 \pi)^2 } \int d^4 x D^2 \bar{D}^2  \( \dua (i \D V) 
     \: \bar{D}^2 \dla (i \D V) \) \ret
& & + {1 \over 4 g^2} \int d^4 x \bar{D}^2 {\tcwb^{\da} \tcwb_{\da} \over 2} 
 - {4 g^2 c^2 \over 2 (8 \pi)^2} \int d^4 x \bar{D}^2 D^2  \( \dbua (i \D V) 
     \: D^2 \dbla (i \D V) \) \ret
& & + {c \over 8 \pi} \int d^4 x d^4 \th 
       { \cw^2 \cwb^2 + (D \phi)^2 (\bar{D} \fb)^2 - 
         \cw^{\da} \dla \f \cwb^{\da} \dbla \fb \over \f^2 \fb^2}  \nn \ ,
\eeqs

\ni
where

\be
 \tilde{\fb} = \fb + {2 g^2 c^2 \over 8 \pi} D^2 \, \D \J  \;\; , \;\;\;
 \tilde{\cw}_{\a} = \cw_{\a} - {4 g^2 c^2 \over 8 \pi} \bar{D}^2 
    \dla (i \D V) \ .
\label{redefi}
\ee

The additional six derivative operators suggest that there are 
acceleration terms (and also part of the $v^6, \cf^6$ terms in 
$S^{(6)}_{eff}$) that enter in the redefinition of the kinetic term. 
We therefore conjecture that all the
acceleration terms $O^{(6)}$ in $S^{(6)}_{eff}$ can be absorbed in 
a redefinition of $S^{(2)}_{eff}$ and $S^{(4)}_{fermi}$. We also 
conjecture that acceleration terms and part of the velocity terms 
in $S^{(n > 6)}_{eff}$ can be additionally absorbed in 
$S^{(4)}_{fermi}$ to make it depend in the same superfields as 
$S^{(2)}_{eff}$

\beqs
 S^{(2)}_{eff} + S^{(4)}_{eff} + O^{(6)} + ... & = & 
  {1 \over 2 g^2} \int d^4 x d^4 \th \: \tfb \tf 
 +{1 \over 4 g^2} \int d^4 x D^2 {\tcw^{\a} \tcw_{\a} \over 2}
 + {1 \over 4 g^2} \int d^4 x \bar{D}^2 { \tcwb^{\da} \tcwb_{\da} \over 2}  
  \ret
& & + {c \over 8 \pi} \int d^4 x d^4 \th 
       { \tcw^2 \tcwb^2 + (D \tf)^2 (\bar{D} \tfb)^2 - 
         \tcw^{\a} \dla \tf \tcwb^{\da} \dbla \tfb \over \tf^2 \tfb^2} \ 
\eeqs

\ni
Note that the bosonic part of this $N=1$ superspace action does not
contain any acceleration terms

\beqs
 S^{bos}_{\tf, \cf} & = &\int d^4 x \( - {1 \over 2 g^2}
  {\pa \tfb \cdot \pa \tf \over 2} - {1 \over 4 g^2} 
 { \tcf^2 + \tcfb^2 \over 2} \ri \\
& & \le \qquad \qquad + {c \over 8 \pi} {\squart \tcf^2 \tcfb^2 +  
 \squart (\pa \tf)^2 (\pa \tfb)^2 - \tcf^{\b \a} \pa_{\a \db} \tf 
 \tcfb^{\db \da} \pa_{\b \da} \tfb \over \tf^2 \tfb^2} + auxiliary \) \ . \nn
\label{bosonic_s4}
\eeqs

\ni 
Identifying 

\be
 F^2 = {1 \over 4 g^2 T_3} \tcf^2 \;, \;\;\;\; 
 X_i X_i = {1 \over g^2 T_3} \tf \tfb \; , \; i=4,5 \;, \;\;\;\;  
 \cn = 2 \pi c  \ ,
\label{identif}
\ee

\ni
we can exactly match (\ref{bosonic_s4}) with the Taylor expansion of   
(\ref{Born_In}) up to four derivatives. In fact, we expect that after
absorbing $O^{(6)}$ in $S^{(2)}_{eff} + S^{(4)}_{fermi}$ the remainder 
of $S^{(6)}_{eff}$ contains four or more fermionic superfields 
({\em i.e.} its bosonic component depends on velocities and field 
strengths but not their derivatives) 

\be
 S^{(6)}_{eff} - O^{(6)} = S^{(6)}_{fermi} \ . 
\ee

\ni
A fraction $\ep$ of the coefficient of this fermionic action is also 
used in the redefinition, 
and the rest of it can absorb acceleration terms from $S^{(2n>6)}_{eff}$ 
to become a functional of $\tf, \tcw_\a$. Matching the bosonic components
of $(1-\ep)S^{(6)}_{fermi} (\tf, \tcw_\a)$ with the six derivative terms of
the DBI action is then a highly nontrivial test of our redefinition.

 If we were to extend our redefinition of the $N=1$ superfields in 
the gauge multiplet to include arbitrary higher derivative pieces, we 
would expect to find a redefined action containing some acceleration 
terms in $S^{(2n>6)}_{eff} (\tf, \tcw_\a)$ \cite{andreev}. Then the $N=4$ 
SYM effective action and the DBI action plus its higher genus 
corrections would agree to all orders.

 When we try to test our conjecture for $S^{(6)}_{eff}$ we find an 
important obstacle: we have mentioned that duality fixes the 
numerical coefficient of the contributions to $S^{(6)}_{eff}$ which 
are analytic in $1/ \t$ (see (\ref{addit_op})). Therefore we can only 
test our conjecture for that particular piece $S^{(6)}_{eff} (1/\t)$. 
The coefficients of contributions proportional to $1/ |\t|$ are not
constrained by duality, and we can only hope that this
comparison will fix them.

 The detailed calculation of the higher order terms $O^{(6)}$ needed
for our redefinition to work is quite tedious and we have included it
in the appendix C. The main result from that calculation is the following:
the only term in $S^{(6)}_{eff} (1/\t)$ which is not absorbed in 
$S^{(2)}_{eff} + S^{(4)}_{eff}$ and contributes to the bosonic gauge 
action is given by 

\beqs
 (1-\ep) S^{(6)}_{fermi} (1/\t) & = & - { 2 g^2 c^2 \over (8 \pi)^2} 
 \int d^4 x \, d^4 \th \; { \shalf \l[ (D \tcw)^2 + (\bar{D} \tcwb)^2 \r] 
        (\tcw)^2 (\tcwb)^2 \over \tf^4 \tfb^4 } \ret
 & = & -{ g^2 c^2 \over 4 (8 \pi)^2} \int d^4 x \;  
    { (\tcf^2 + \tcfb^2) \tcf^2 \tcfb^2 \over \tf^4 \tfb^4 } + fermions + 
     auxiliary\ . 
\eeqs

 Our identification (\ref{identif}) provides again an exact match to
the six derivative term in the Taylor expansion of the gauge DBI 
action.

\beqs
 S_{BI} & = & \int d^4 x T_3 h^{-1} \( 1 - \sqrt{ 1 + h 
  (F^2 + \bar{F}^2) + \squart h^{2} (F^2 - \bar{F}^2)^2 } \) 
                   \label{BI_exp}\\
& = & \int d^4 x \( - T_3 {F^2 + \bar{F}^2 \over 2} + 
 \shalf T_3 h F^2 \bar{F}^2 - \squart T_3 h^{2} 
       F^2 \bar{F}^2 (F^2 + \bar{F}^2) + \dots \) \nn \\
& = & \int d^4 x \( - T_3 {F^2 + \bar{F}^2 \over 2} + 
 4 \cn {F^2 \bar{F}^2 \over \rt^4} - {16 (\cn)^2 \over T_3}  
 {F^2 \bar{F}^2 (F^2 + \bar{F}^2) \over \rt^8} + \dots \) \nn \ ,
\eeqs

\ni
where the DBI field strength is written with $SL(2,C)$ spinor indices
and we have defined $\rt^2 = 4 \pi X^i X^i = 4 \pi \tf \tfb / g^2 T_3$. 
Note that without the redefinition, setting acceleration terms to zero
in $S^{(6)}$ would not be enough to achieve the agreement between both
descriptions of the 3-brane dynamics.

 To try to gain some understanding about the physical meaning of our  
redefinition we consider again the duality transformation of the gauge
scalars: in the brane picture, the DBI action in a curved background is 
self-dual \cite{tseyt}, but the transversal separation $\rt^2$ 
remains unchanged \cite{aga, park_rey} when we dualize the $N=0$ 
gauge field strength of the DBI action
 
\beqs
 S^D_{BI} & \!\! = & \!\!\! \int d^4 x \: T_3 \, h^{-1}(\rt) 
 \( 1 - \sqrt{ 1 + h (\rt) (F^2_D + \bar{F}^2_D) 
  + \squart h^{2}(\rt) (F^2_D - \bar{F}^2_D)^2} \: \) \\ 
&\!\! = & \!\!\! \int \! d^4 x \: T_3 \, h^{-1}(\rt) \( 1 - \sqrt{ 1 + h (\rt) 
  (F^2 + \bar{F}^2) + \squart h^{2}(\rt) (F^2 - \bar{F}^2)^2 } +
   i F \cdot F_D - i \bar{F} \cdot \bar{F}_D \) \nn   
\eeqs

\ni
whereas in the SYM picture the scalar $r^2 = 4 \pi \f \fb / g^2 T_3$ 
transforms in accordance with

\beqs
 r_D^2 = {4 \pi \over g_D^2 T_3} W_D \bW_D | & = & \le {g^2 \over 4 \pi T_3} 
 \( { 4 \pi \over g^2} W + 2 c {\bd4 \ln \bW \over W} + ...\) 
 \( { 4 \pi \over g^2} \bW + 2 c {D^4 \ln W \over \bW} + ...\) \ra| \nn \\
& \simeq & r^2 \( 1 + {4 \pi c \over g^2 T_3^2} {\bar{\cf}^2 \over r^4} +...\) 
  \( 1 + {4 \pi c \over g^2 T_3^2} {\cf^2 \over r^4} +...\) \ .
\eeqs
 
\ni
The self-dual scalar $\rt^2$ is therefore a nonlinear combination of
$r^2$ and the gauge field strengths which remains invariant
under duality (we are ignoring the velocity terms). Up to two 
derivatives it is given by  

\beqs
 \rt^2 & = & {4 \pi \over g^2 T_3} \le W \(1 + { c g^2 \over 4 \pi} 
 {\bd4 \ln \bW \over W^2} + ...\) \; \bW \(1 + {c g^2 \over 4 \pi} 
 {D^4 \ln W \over \bW^2} + ...\) \ra| \label{redef_scalar} \ret \ret
& = & r^2 \( 1 + {2 \pi c \over g^2 T_3^2} {\bar{\cf}^2 \over r^4} \)
 \( 1 + {2 \pi c \over g^2 T_3^2} { \cf^2 \over r^4} \) + ...
 = { r^2 + r_D^2 \over 2} + ... 
\eeqs

\ni
Remarkably, this is consistent with the shift (\ref{prepot}, 
\ref{redefi}) needed to remove the acceleration terms from the
effective action. 
For a constant gauge field strength we can see that in the SYM picture
the separation $r$ between the probe 3-brane and the source 
3-brane is normalized by a scale that contains information about the 
charge density in the 3-brane ({\em i.e.} about the internal energy 
density)

\be
 \rt^4 = r^4 + {4 \pi c \over g^2 T_3^2} (\cf^2 + \bar{\cf}^2) + ...
\ee

\ni
Perhaps this rescaling accounts for a correction to the background 
metric induced by the gauge fields living in the 3-brane. In the 
probe-source picture this information is implicit in the 
separation variable. 

 In summary, we are able to successfully compare the effective action
of $N=4$ SYM with the DBI action of a gauge field strength 
living in the D3-brane up to terms with six space-time
derivatives. This comparison reveals the surprising fact that the 
gauge scalar and the separation of the probe and source 3-branes are 
not simply proportional to each other. Contrary to common lore, we
find that the relation involves a nonlinear function of the field 
strengths living in the probe. We emphasize that our results are valid
for arbitrary $g^2$ and arbitrary $N_c$. They do not
contradict previous ideas \cite{mald_conj} \cite{douglas_taylor}, but 
merely refine them: 
the issue of which variable $r$ transforms linearly under $N=2$ SUSY 
has not been addressed in comparing the DBI action with the SYM
effective action.

We have restricted our analysis to the abelian sector of $N=4$ SYM 
theory in four dimensions. It seems plausible that in the unbroken 
gauge theory of overlapping 3-branes, an extension of the redefinition 
we have presented is also necessary. This would have important 
consequences for the identification of the fields appearing in the 
correlators of the superconformal four-dimensional theory that lives 
in the boundary of $AdS_5$ \cite{trivedi}. Such fields would not   
correspond exactly to the asymptotic states of the $N=4$ SYM theory    
where the supersymmetry is linearly realized.

\section{Conclusions and open problems} 

 In this article we have succeeded in constructing the momentum
expansion of a self-dual $N=4$ SYM effective action, which by 
scale and $U(1)_R$ invariance turns out to be also an expansion on the
inverse of the holomorphic gauge coupling $\t = {\th_v \over 2 \pi} + 
i {4 \pi \over g^2}$ (and its absolute value if we include more general 
contributions).

Our most striking observation is the redefinition of the
Higgs field needed to reconcile $N=2$ supersymmetry and the
self-duality of the DBI action.

 The expansion of the effective action that we have found is
analytic in $g^2$. For $\th_v\neq 0$, the terms that we find  
seem to be in conflict with the type of contributions obtained from 
instantons. We find the apparent contradiction that there are 
nonvanishing instanton contributions to $S^{(6)}_{eff}$ (and to higher
order terms) \cite{diego} that introduce a dependence of the effective
action on $\th_v$, and yet are not analytic in $g^2$.

Since our analysis only specifies the coefficients of some terms but
not others, it is possible that these terms can be completed to have
the appropriate $\th_v$ dependence; however, we do not see how to do
this. On the other hand, our results could 
signal a breakdown in the instanton series. This seems unlikely in
light of the results of \cite{bagr}.

\section*{Acknowledgments}

 We are very grateful to R. von Unge, D. Bellisai and Y. Chepelev for 
useful exchanges and interesting suggestions. We would also like to thank 
T. Banks, A. Goldhaber, H. Nastase and G. Shiu for discussions. FGR would like
to thank the Les Houches summer school ``Field Theory: Perspectives and
Prospectives'' June 1998, where part of this research was presented in
one of the participants talks. This research was 
partially supported under NSF grant No. Phy 9722101.

\section*{\LARGE\bf Appendix A}

 In this appendix we generalize the results of \cite{GR} to an arbitrary 
$SU(N_c)$ gauge group spontaneously broken to $SU(N_c-1) \times U(1)$
or smaller subgroups. In 
that reference we found the nonholomorphic potential in $N=2$ superspace 
of a generic $N=4$ SYM theory as a momentum integral

\be
 \ch (W, \bar{W}) = {1 \over 2} \int {d p^2 \over (4 \pi)^2 
 p^2} \, \tr_A \ln \( 1 + {W \bar{W} + \bar{W} W \over 2 p^2} \) \ , 
\label{n2_result}
\ee

\ni
and we were able to evaluate this integral for the abelian subgroup of 
$SU(2)$. The key to generalize this result is to evaluate the gauge 
operator $W \bW + \bW W$ for a particular $U(1)$ subgroup of $SU(N_c)$.
This is very straightforward when we write a generator in the adjoint 
representation in terms of its fundamental representation 

\be
 \( T^a_A \)^{i \ \ k}_{\ j;\ \ l} = \( T^a_F \)^i_l \d^k_j - 
                                   \( T^a_F \)^k_j \d^i_l  \ .
\ee

\ni
We want to break $SU(N_c)$ by giving a nonzero {\em v.e.v.} to the 
gauge scalar aligned with a particular linear combination of Cartan generators 
$W^{i \ \ k}_{\ j,\ \ l} = \langle W_a \rangle 
\( T^a_A \)^{i \ \ k}_{\ j;\ \ l}$. The operator we want to evaluate is 
therefore

\be
 \(W \bW + \bW W \)^{i \ \ k}_{\ j;\ \ l} = (W \bW + \bW W)^k_{\ j} \d^i_{\ l}
 + (\bW W + W \bW)^i_{\ l} \d^k_{\ j} - 2 W^i_{\ l} \bW^k_{\ j} 
 - 2 W^k_{\ j} \bW^i_{\ l} \ .
\label{gauge_op}
\ee

 Once we choose a linear combination in the fundamental representation, 
the corresponding gauge scalars parameterize the transversal 
positions of the associated $N_c$ 3-branes. We are mostly interested 
on the configuration
of $N_c-1$ overlapping 3-branes separated from another 3-brane. The 
$U(1)$ gauge scalar has a {\em v.e.v.} 

\be
 W = \langle W \rangle T^{N_c-1}_F = {1 \over \sqrt{2 N_c(N_c-1)} }
       diag\(w,w,...,w,(1-N_c)w\) \ .
\ee

\ni
Substituting this background field in (\ref{gauge_op}) we find a 
diagonal adjoint representation matrix $(W \bW + \bW W)_A$ with $2(N_c-1)$ 
nonzero elements

\be
 \shalf (W \bar{W} + \bar{W} W)_A = (W \bar{W})_A = (\bar{W} W)_A =
 \( \begin{array}{ccccccc}
           0 & . & . & . & . & . & . \\
           . & . & . & . & . & . & . \\
           . & . & {N_c \over 2 (N_c-1)} w \bar{w} & . & . & . & . \\ 
           . & . & . & . & . & . & . \\  
           . & . & . & . & {N_c \over 2 (N_c-1)} w \bar{w} & . & . \\
           . & . & . & . & . & . & 0 \\ 
       \end{array} \) \ .
\label{eigen_suN}
\ee

\ni
This facilitates the evaluation of the trace in (\ref{n2_result})

\be
 \ch (W, \bar{W}) = {N_c-1 \over (4 \pi)^2} \int {d p^2 \over p^2} 
 \; \ln \( 1 + {N_c \over 2 (N_c-1)} {W \bW \over p^2} \) \ .
\label{h-1loop}
\ee

 Using dimensional regularization as in \cite{GR} we are able to
perform the momentum integral and we obtain

\beqs
 \meas \; \ch^{Abel} (W, \bar{W}) & = & {N_c-1 \over (4 \pi)^2 } \cmeas
 \( \ln {N_c \over 2 (N_c-1)} {W \bar{W} \over \mu^2 } \)^2 \ret
 & = & {N_c-1 \over (4 \pi)^2 } \cmeas \; \ln W \ln \bW \ .
\eeqs

 If we try to perform a a similar calculation for the massless 
nonabelian multiplets associated with the unbroken $SU(N_c-1)$ theory
we find IR divergences that need to be regulated. We therefore simplify
our analysis by focusing on the $U(1)$ sector.

Notice that the numerical factor in each diagonal matrix element of 
(\ref{eigen_suN}) only changes the normalization scale in the final
expression, and since we have a full $N=2$ superspace measure the
action is scale independent. 

This observation is important when we want to consider different
3-brane configurations parameterized by other gauge scalar {\em
v.e.v.}'s which are diagonal in the fundamental representation, and
break $SU(N_c)$ further. We
obtain again a diagonal operator $(W \bW)_A$ in the adjoint
representation, with different coefficients in its nonzero elements. 
Such coefficients end up modifying the normalization scale, and they 
drop out of the nonholomorphic potential. The only relevant quantity 
is the overall number of nonzero eigenvalues in $(W \bW)_A$. 
It is straightforward to see that these eigenvalues are given by the 
differences of eigenvalues of $(W \bW)_F$ in the fundamental 
representation \cite{chep}. Hence, a generically broken $SU(N_c)$
configuration\footnote{Note added in proof: a derivation of the same
$N=2$ nonholomorphic potential corresponding to this configuration 
appeared in \cite{kuz2,rik_lowe} shortly after this manuscript was 
completed.} 

\be
 W = \langle W_a \rangle T^a_F = diag\(w_{11},w_{22},...,w_{N_c N_c}\) 
  \;\;\; , \;\;\;\; \sum_{i=1}^{N_c} w_{ii} = 0 \ ,
\ee

\ni
will give a nonholomorphic potential (if $W_{ii}=W_{jj}$ the 
corresponding term is absent from the sum)

\be
 \sum_{i<j} S^{(4)}_{ij} = { 1 \over (4 \pi)^2 } \meas \sum_{i<j} 
    \ln \( W_{ii} - W_{jj} \) \ln \( \bW_{ii}- \bW_{jj} \) \ .
\ee

In these more general configurations, the correspondence with the DBI 
action is more subtle. Since the tree level action of these
abelian components can also be decomposed for an $SU(N_c)$ group

\be
 S^{(2)} = {\car e \over 2 g^2} \int d^4 x D^2 Q^2 \sum_{i=1}^{N_c} W_{ii}^2 = 
  {\car e \over 2 g^2 } \int d^4 x D^2 Q^2 {1 \over 2 N_c}
    \sum_{i=1}^{N_c} \sum_{j=1}^{N_c} (W_{ii} - W_{jj})^2 = 
   \sum_{i<j} S^{(2)}_{ij}
\ee

\ni
the interactions of the 3-branes seem to group pairwise 
\cite{douglas_taylor}. We would have
a probe-source DBI description for each pair $i,j$ of 3-branes, 
whose Taylor expansion should be matched (up to four spacetime
derivatives) with an effective action $S^{(2)}_{ij} + S^{(4)}_{ij}$.

\section*{\LARGE\bf Appendix B}
    
 In this appendix, we will simply quote the higher order result. We
give the self-dual gauge effective action that depends on scale and $U(1)_R$ 
invariant chiral variables up to terms with sixteen space-time 
derivatives. Then we present the gauge effective action that depends
on $\cd$ and $\cp^{ab}$ up to eight spacetime derivatives. 
Since the expression is rather long, the following notation is convenient

\be
 \cd \equiv {D^4 \over \bt \bW^2} , \;\;\;\;\;\; 
 B \equiv {\bd4 \ln \bW \over \t W^2} , \;\;\;\;\;\;
 \int \equiv \cmeas \ .
\ee 

\ni 
Using this definition, the effective action is 

\beqs
 8 \pi S_{eff} & = & 8 \pi \( S^{(2)} + S^{(4)} + S^{(6)} + S^{(8)} 
  + S^{(10)} + S^{(12)} + S^{(14)} + S^{(16)} + \cdots \) \ret
 & = & \ci m \icmeas \; \shalf \t W^2 + c \int \: ln W \,\ ln \bW \ret
 & & \l[ + \; i c^2 \int \, \shalf \,ln \bW B \ri \label{16act}  \\
 & & + \; c^3 \int \( \half \k^{(8)}_1 \, \, B \bar{B} 
                        + \k^{(8)}_2 \, \ln \bW \, B^2 \) \ret
 & & + \; i c^4 \int \( \k^{(10)}_1 \, \ln \bW \, B^3 
          + \k^{(10)}_2 \, B^2 \bar{B} + \k^{(10)}_3 \,B \,\cd (B) \) \ret
 & & + \; c^5 \int \( \k^{(12)}_1 \, \ln \bW \, B^4 
       + \k^{(12)}_2 \, B^3 \bar{B} + \shalf \k^{(12)}_3 B^2 {\bar{B}}^2 
       + \shalf \k^{(12)}_4 \cd (B) \cdb (\bar{B})   \ri \ret
 & &  \le \qquad \qquad + \: \k^{(12)}_5 \, B^2 \, \cd (B)
        + \k^{(12)}_6 \,B \bar{B} \, \cdb (\bar{B}) \)  \ret
 & & + \; i c^6 \int \( \k^{(14)}_1 \ln \bW \, B^5 + \k^{(14)}_2 B^4 \bar{B}
      + \k^{(14)}_3 B^3 {\bar{B}}^2 + \k^{(14)}_4 B^3 \, \cd (B) \ri \ret
 & & \qquad \qquad + \; \k^{(14)}_5 B^2 \, \cd (B^2)      
     + \k^{(14)}_6 B (\cd (B))^2 + \k^{(14)}_7 B^2 \bar{B} \, \cd (B) \ret
 & & \qquad \qquad + \; \k^{(14)}_8 B \bar{B}^2 \cd (B)      
      + \k^{(14)}_9 B \, \cd (B) \, \cdb (\bar{B})
      + \k^{(14)}_{10} \cdb (\bar{B}) \, \cd (B^2) \ret
 & & \qquad \qquad \le + \k^{(14)}_{11} \cd (B) \, \cdb (\cd (B)) \)\nn\\
 & & + \; c^7 \int \( \k^{(16)}_1 \ln \bW B^6 + \k^{(16)}_2 B^5 \bar{B} 
        + \k^{(16)}_3 B^4 \bar{B}^2 + \shalf \k^{(16)}_4 B^3 \bar{B}^3
        + \k^{(16)}_5 B^4 \cd (B)    \ri   \ret
 & & \qquad \qquad + \; \k^{(16)}_6 B^3 \bar{B} \cd (B)                
     + \k^{(16)}_7 B^3 \bar{B} \cdb (\Bb) + \k^{(16)}_8 B^3 \cd (B^2)  
     + \k^{(16)}_9 B^2 \cd (B)^2  \ret
 & & \qquad \qquad + \; \k^{(16)}_{10} B^2 \Bb \cd (B^2)
       + \k^{(16)}_{11} B^2 \cd (B) \cdb (\Bb)          
       + \k^{(16)}_{12} B^2 \Bb^2 \cd (B) \ret
 & & \qquad \qquad + \; \k^{(16)}_{13}  B \Bb \cd (B)^2
       + \k^{(16)}_{14} B \cd (B) \cdb (\Bb^2)
       + \shalf \k^{(16)}_{15} B \Bb \cd (B) \cdb (\Bb) \ret           
 & & \qquad \qquad + \; \k^{(16)}_{16} B \cdb (\Bb) \cd (B^2)
       + \k^{(16)}_{17} B \cdb (\Bb) \cd (\cdb (\Bb))        
       + \k^{(16)}_{18} B \cd (B) \cdb (\cd (B))  \ret
 & & \qquad \qquad + \; \k^{(16)}_{19} \cd (B)^2 \cdb (\Bb)
       + \k^{(16)}_{20} \cd (B^3) \cdb (\Bb) 
       + \k^{(16)}_{21} \cd (B^2) \cdb (\D(B)) \ret                 
 & & \le \le \qquad \qquad + \; \shalf \k^{(16)}_{22} \cd (B^2) \cdb
    (\Bb^2) \) 
     \;\;\;\;\;\; + \;\; \dots \;\;\;\;\;\;\; +c.c. \;\;\;\; \r]  \ .
\eeqs 

\noindent 
where $\k^{(8)}_1, \k^{(10)}_3, \k^{(12)}_4, \k^{(16)}_4, \k^{(16)}_{15}$ 
and $\k^{(16)}_{22}$ are real and

\beqs
 \k^{(10)}_1 & \equiv & - 4 \k^{(8)}_2 + {7 \over 6} \ret
 \k^{(10)}_2 & \equiv & - 2 \k^{(8)}_1 - 3 \k^{(8)}_2 + 2 \ret
 \k^{(10)}_3 & \equiv & 2 \k^{(8)}_1 - 1  \\
 \k^{(14)}_1 & \equiv & - 8 \k^{(12)}_1 - 72 \k^{(8)}_2 +8 {\k^{(8)}_2}^2
 + {121 \over 5} \ret
 \k^{(14)}_2 & \equiv & - 5 \k^{(12)}_1 - 20 \k^{(8)}_1 - 6 \k^{(12)}_2
   - 104 \k^{(8)}_2 + 8 \k^{(8)}_1 \k^{(8)}_2 + 48  \ret
 \k^{(14)}_3 & \equiv & 2 \k^{(12)}_2 + 16 \k^{(8)}_1 + 32 \k^{(8)}_2 
   + 14 \bar{\k}^{(8)}_2 - 4 \k^{(12)}_3 - 12 \k^{(8)}_2 \bar{\k}^{(8)}_2 
   - {76 \over 3}    \ret
 \k^{(14)}_4 & \equiv & \k^{(12)}_2 + 16 \k^{(8)}_2 
   + {38 \over 3} \k^{(8)}_1 - 4 \k^{(12)}_5
   - 4 \k^{(8)}_1 \k^{(8)}_2 - {38 \over 3}          \ret 
 \k^{(14)}_5 & \equiv & 4 \k^{(8)}_1 + 6 \k^{(8)}_2 - 2 \k^{(12)}_5
   - {9 \over 2} {\k^{(8)}_2}^2 - 4          \ret
 \k^{(14)}_6 & \equiv & 2 \k^{(8)}_1 + \bar{\k}^{(12)}_6 
   + 2 \bar{\k}^{(8)}_2 - 2           \ret
 \k^{(14)}_7 & \equiv & - 4 \bar{\k}^{(12)}_6 - 20 \k^{(8)}_1 
   + 2 \k^{(12)}_5 - 12 \k^{(8)}_2 - 12 \bar{\k}^{(8)}_2 
   + 6 \k^{(8)}_1 \k^{(8)}_2 + 2 \k^{(12)}_3 + 20 \ret
 \k^{(14)}_8 & \equiv & 3 \bar{\k}^{(12)}_2 + 20 \k^{(8)}_1 
   + 4 \bar{\k}^{(12)}_6 - 2 {\k^{(8)}_1}^2 + 36 \bar{\k}^{(8)}_2 - 25 \ret
 \k^{(14)}_9 & \equiv & 14 \k^{(8)}_1 + 6 \k^{(8)}_2 + 2 \k^{(12)}_6 
   - 2 \k^{(12)}_5 - 2 \k^{(12)}_4 -2 {\k^{(8)}_1}^2 - 10       \ret
 \k^{(14)}_{10} & \equiv & 6 \k^{(8)}_1 + 3 \k^{(8)}_2 
   - 3 \k^{(8)}_1 \k^{(8)}_2 - \k^{(12)}_5 - 2 \k^{(12)}_4 - 4   \ret
 \k^{(14)}_{11} & \equiv & - 2 \k^{(8)}_1 + \k^{(12)}_4 
   + {1 \over 2} {\k^{(8)}_1}^2 + 1        \nn
\eeqs

 In section 3, we pointed out that self-duality
does not exclude operators such as
(\ref{simp_ops}). We also introduced a simplified notation for them

\be
 \cp^{ab} = {\bar{D}^{2 a b} \over \sqrt{\t} W } \;\;\; \ , \;\;\;\;\; 
 E^{ab} =  {\bar{D}^{2 a b} \ln \bW \over \sqrt{\t} W }
    =  {\bar{D}^{2 a b} \ln (\sqrt{\t} \bW) \over \sqrt{\t} W } \ .
\ee

\ni
It is possible to construct a self-dual effective action  
that includes these operators     

\beqs
 8 \pi \: S_{eff} & = & 8 \pi \( S^{(2)} + S^{(4)} + S^{(6)} 
                     + S^{(8)} + S^{(10)} + \dots \) \nn\\
 & = & {\ci}m \( \icmeas \; \shalf \t W^2 \) 
                         + \: c \meas \: \ln W \, \ln \bW           \ret
 & &  \left[ \;\; + \: c^2  \int \( \; i \shalf  B \: \ln \bW \;\;
         + \: \shalf \lambda^{(6)} \; E\cdot \bar{E} \) \ri \ret
 & &  \;\; + \: c^3 \int \(  \shalf \k^{(8)}_1 \, B \: \bar{B}     
      \; + i \lambda^{(6)} \, B ( \cp \cdot E - \bar{E} \cdot E) \; 
         + \: \k^{(8)}_2 \; B^2 \ln \bW \; \)  \ret 
 & &  \;\; + \: c^4 \int \( \(-8 i \k^{(8)}_1 + i {7 \over 3} \)   \; 
            B^3 \, \ln \bW + \(2 i \k^{(8)}_2 - i\) \; B \cd (B) \ri \ret    
 & & \qquad \qquad + \: \(-6 i \k^{(8)}_1 - 4 i\k^{(8)}_2 + 4 i\) \; 
              B^2 \, \bar{B}   \ret 
 & & \qquad \qquad + \: {\lambda^{(10)}_1 } \; B^2 
   ( \bar{E} \cdot E - \cp \cdot E ) \; + \shalf \lambda^{(10)}_2 \; B \bar{B}
           \(\bar{E} \cdot E - \cp \cdot E - \cpb \cdot (\bar{E}) \)  \ret 
 & & \qquad \qquad + \: {\lambda^{(10)}_3  } \; B E\cdot \cp (B) \;    
        + \lambda^{(10)}_4 \; \bar{E} \cdot E \cd (B) \;   
        + \lambda^{(10)}_5 \; \cpb \cdot (\bar{E}) \cd (B)   \ret 
 & &  \qquad \qquad + \: \shalf \lambda^{(10)}_6 \; \cpb \Bb \cdot \cp (B) \ret
 & & \le \le \qquad \qquad - \: {i \over 2}  {\( \lambda^{(6)} \)^2}  
         \( \bar{E} \cdot E - \cp \cdot E \) \cdb 
         \( \bar{E} \cdot E - \cp \cdot E \) + \dots  \) +c.c. \;\;\; \r] \nn
\eeqs 
         
 Note that we could have also added a term 
$c^3 \lambda^{(8)} (\bar{E} \cdot E)^2$ to $S^{(8)}_{eff}$ which would be 
mapped to itself at lowest order under duality, preserving the
self-duality at this order.       

\section*{\LARGE\bf Appendix C}    

In this appendix we present the detailed calculation of the $N=1$
components with six space-time derivatives which are needed in the SYM
effective action for our redefinition (\ref{redefi}) to work. Let us 
begin our analysis identifying the six derivative operators that are 
implied by the redefinition of two and four derivative operators
(for brevity we write $V$ and $\J$ instead of $\D V$ and $\D \J$)

\beqs
 O^{(6)} & = &  \int d^4 x \, d^4 \th \; {2 g^2 c^2 \over (8 \pi)^2} \( 
   (D^2 \J) (\bar{D}^2 \bar{\J}) - \l[ \dua (i V) \bar{D}^2 \dla (i V) \r] 
 - \l[ \dbua (i V) D^2 \dbla (i V) \r] \) \ret
& & \qquad + {c \over 8 \pi} { \l[ {-4 g^2 c \over 8 \pi} \bar{D}^2
   \dua (i V) \r] \cw_{\a} (\cwb)^2 + (\cw)^2 \l[ {-4 g^2 c \over 8 \pi} 
   D^2 \dbua (i V) \r] \cwb_{\da} \over \f^2 \fb^2 } \\
& & \qquad + {c \over 8 \pi} { \dua \f \l[ {2 g^2 c \over 8 \pi} 
   \dla \bar{D}^2 \bar{\J} \r] (\bar{D} \fb)^2  
 + (D \f)^2 \dbua \fb \l[ {2 g^2 c \over 8 \pi}
    \dbla D^2 \J \r] \over \f^2 \fb^2 } \ret
& & \qquad - {c \over 8 \pi} { \l[ {-4 g^2 c \over 8 \pi} \bar{D}^2
   \dua (i V) \r] \dla \f \; \cwb^{\da} \dbla \fb 
 + \cw^{\a} \l[ {2 g^2 c \over 8 \pi} \dla \bar{D}^2 \bar{\J} \r] 
   \cwb^{\da} \dbla \fb \over \f^2 \fb^2 } \ret
& & \qquad - {c \over 8 \pi} { \cw^{\a} \dla \f 
 \l[{-4 g^2 c \over 8 \pi} D^2 \dbua (i V) \r] \dbla \fb  
 + \cw^{\a} \dla \f \; \cwb^{\da} 
   \l[ {2 g^2 c \over 8 \pi} \dbla D^2 \J \r] \over \f^2 \fb^2 } \ret
& & \qquad - {c \over 8 \pi}  
       { \cw^2 \cwb^2 + (D \phi)^2 (\bar{D} \fb)^2 - 
         \cw^{\a} \dla \f \; \cwb^{\da} \dbla \fb \over \f^2 \fb^2} \;
   {4 g^2 c \over 8 \pi} \( {\bar{D}^2 \bar{\J} \over \f} + 
 {D^2 \J \over \fb} \)   \nn \ .
\eeqs

 To simplify our analysis, we ignore all terms containing the
auxiliary superfields $D^2 \f$, $\dla \cw^{\a}$ and their conjugates. 
The first type can be absorbed as a higher order correction in the
redefinition of the scalar kinetic term $\fb \f$ and the second can be
absorbed in a redefinition of the gauge spinor kinetic term $ \cw^2 +
c.c.$. Such higher order corrections will in turn require
acceleration terms from $S^{(8)}_{eff}$ and $S^{(10)}_{eff}$. We 
know the general form of these higher contributions but not their precise
numerical coefficients, and therefore we can only establish that the
required acceleration terms are generically present in the effective 
action\footnote{It is worth mentioning that the elimination of auxiliary
superfields through redefinition of the physical ones is a feature we 
have encountered in the study of the supersymmetric formulation of
the BI action \cite{rocek_tseyt} and the 3-brane living in six dimensions
\cite{nonlin}.}.

 After integrating by parts and some algebra we find that it is useful
to classify the terms without auxiliary superfields by counting the
number of fermionic superfields

\beqs
 O^{(6)}_{0} & \!\! = \!\! & {g^2 c^2 \over (8 \pi)^2} \int d^4 x \, d^4 \th 
 \( 2 { \shalf (D \cw)^2 \: \shalf (\bar{D} \cwb)^2 \over \f^3 \fb^3}
 + 2 { \shalf (\bar{D} D \f)^2 \: \shalf (D\bar{D} \fb)^2 \over \f^3 \fb^3}
 + 3 { \shalf (D \cw)^2 \: \shalf (\bar{D} D \f)^2 \over \f^4 \fb^2}
  \ri \ret
& & \le \qquad \qquad + 3 { \shalf (\bar{D} \cwb)^2 \; \shalf (D\bar{D} \fb)^2 
  \over \f^2 \fb^4} - {(\dua \cw^{\b}) (\dbla D_{\b} \f) (\dbua \cwb^{\db}) 
    (\dla \bar{D}_{\db} \fb) \over \f^3 \fb^3} \) 
\eeqs

\beqs
 \lefteqn{ O^{(6)}_{2} = {g^2 c^2 \over (8 \pi)^2} \int d^4 x \, d^4 \th 
   \;\;\;\; 6 {\shalf (\ddbfb)^2 (\dbua \cwb^{\db}) 
           \cwb_{\db} \dbla \fb \over \f^2 \fb^5}
 + 6 {\shalf (D \cw)^2 (\dbua \cwb^{\db}) 
         \cwb_{\db} \dbla \fb \over \f^3 \fb^4} }\ret
& & \qquad + 4 {\shalf (\bar{D} \cwb)^2 (\dua \dbua \fb) \cwb_{\da} \cw_\a 
                     \over  \f^3 \fb^4}
 + 2 {\shalf (\dbdf)^2 (\dua \dbua \fb) \cwb_{\da} \cw_\a \over \f^4 \fb^3}
    \ret
& & \qquad + { (\dbua \cwb^{\db}) (\dbla D_\b \f) (D^\b \bar{D}^{\dg} \fb) 
          \cwb_{\dg} \bar{D}_{\db} \fb \over \f^3 \fb^4}
 - 2 { (\dbla \cwb_{\db}) (\dbua \dua \f) (D^\g \bar{D}^{\db} \fb) 
          \cw_{(\g} D_{\a)} \f \over \f^4 \fb^3 }   \ret
& & \qquad + {1 \over 2} { (\dbua \cwb^{\db}) (\dbla \dla \f) 
   (\dua \bar{D}_{\db} \fb) \cwb^{\dg} \bar{D}_{\dg} \fb \over \f^3 \fb^4} 
 - {1 \over 2} { (\dbua \cwb^{\db}) (\dbla \dla \f) (\dua \bar{D}_{\db} \fb) 
           \cw^{\g} D_{\g} \f \over \f^4 \fb^3}          \ret
& & \qquad + {1 \over 2} { (\dbua \cwb^{\db}) (\dua \bar{D}^{\db} \fb)
  ( i \pa_{\a \dg} \dbla \fb) \cwb^{\dg} \over \f^2 \fb^4}
 + {1 \over 2} { (\dbua \cwb^{\db}) (\dua \bar{D}^{\db} \fb)
   ( i \pa_{\g \da} \dla \f) \cw^{\g} \over \f^3 \fb^3} \ret
& & \qquad - {1 \over 2} { (\dbua \cwb^{\db}) (\dua \bar{D}_{\db} \fb)
   ( i \pa_{\a \da} \cwb_{\dg}) \bar{D}^{\dg} \fb \over \f^2 \fb^4}
 - {1 \over 2} { (\dbua \cwb^{\db}) (\dua \bar{D}_{\db} \fb)
   ( i \pa_{\a \da} \cw_{\g}) D^{\g} \f \over \f^3 \fb^3} \ret 
& & \qquad \qquad \qquad \qquad  + \;\;\; c.c. 
\eeqs

\beqs
 O^{(6)}_4 & = & {g^2 c^2 \over (8 \pi)^2} \int d^4 x \, d^4 \th 
  - 20 { \shalf (\ddbfb)^2 (\cwb)^2 (\bar{D} \fb)^2 \over \f^2 \fb^6 } 
  - 20 { \shalf (D \cw)^2 (\cwb)^2 (\bar{D} \fb)^2 \over \f^3 \fb^5 } \ret
& & \qquad - 3 { \shalf (\ddbfb)^2 (\cwb)^2 (\bar{D} \fb)^2 \over \f^4 \fb^4} 
 + { (\dbdf \cdot \ddbfb) (\cwb)^2 (\bar{D} \fb)^2 \over \f^3 \fb^5} \ret
& & \qquad - 4 { \shalf (\bar{D} \cwb)^2 (\cwb)^2 (\cw)^2 \over \f^4 \fb^4 }
 - 4 { \shalf (\dbdf)^2 (\cwb)^2 (\cw)^2 \over \f^5 \fb^3 }   \ret
& & \qquad + 2 { \shalf (\bar{D} \cwb)^2 (\bar{D} \fb)^2 (D \f)^2 \over 
                  \f^4 \fb^4 }
 + 0 { \shalf (\dbdf)^2 (\bar{D} \fb)^2 (D \f)^2 \over \f^5 \fb^3 }  \ret
& & \qquad - { (\dbdf \cdot \ddbfb) (\bar{D} \fb)^2 (D \f)^2 \over
                  \f^4 \fb^4 } \\
& & \qquad - 2 { \shalf (\bar{D} \cwb)^2 \cwb^{\dg} \bar{D}_{\dg} \fb
    \cw^{\g} D_{\g} \f \over \f^4 \fb^4 }
 + { \shalf (\ddbfb)^2 \cwb^{\dg} \bar{D}_{\dg} \fb \cw^{\g} D_{\g} \f 
       \over \f^3 \fb^5 }  \ret
& & \qquad + 8 { (\dbua \cwb^{\db}) (\dua \bar{D}_{\db} \fb) 
       \cw_{\a} \dbla \fb (\cwb)^2 \over \f^3 \fb^5 }    \ret 
& & \qquad - { (\dbua \cwb^{\db}) (\dbla \dua \f) 
             \cw_{\a} \bar{D}_{\db} \fb (D \f)^2 \over \f^5 \fb^3 }
 - {1 \over 2} { (\dbua \cwb^{\db}) (\dbla \dua \f) 
             \cwb_{\db} \dla \f (\bar{D} \fb)^2 \over \f^4 \fb^4}  \ret 
& & \qquad - { (\dbua \cwb^{\db}) (\dua \dbla \fb) 
             \cwb_{\db} \dla \f (\bar{D} \fb)^2 \over \f^3 \fb^5}  
 + {1 \over 2} { (\dbua \cwb^{\db}) (\dua \dbla \fb) 
             \cw_{\a} \bar{D}_{\db} \fb (D \f)^2 \over \f^4 \fb^4} \ret
& & \qquad + 8 { (\dbua \cwb^{\db}) (\dua \cw^{\b}) 
             \cwb_{\db} \dbla \fb \cw_{\b} \dla \f \over \f^4 \fb^4} 
 + 0 { (\dbua D^{\b} \f) (\dua \bar{D}^{\db} \fb) 
             \cwb_{\db} \dbla \fb \cw_{\b} \dla \f \over \f^4 \fb^4} \ret
& & \qquad + {1 \over 4} { (\pa \bar{D} \fb)^2 (\cwb)^2 \over \f^2 \fb^4}
 + {1 \over 4} { (\pa \cw)^2 (D \f)^2 \over \f^4 \fb^2 }  
 + {1 \over 2} { (\pa_{\b \db} \dbla \fb) (\pa^{\b \db} \cw^\a) 
                 \cwb^{\da} \dla \fb \over \f^3 \fb^3}     \ret
& & \qquad - { (\dua \dbua \fb) (i \pa_{\a \db} \dbla \fb) 
                  \bar{D}^{\db} \fb (\cwb)^2 \over \f^2 \fb^5 } 
 - {1 \over 2} { (\dbua \cwb^{\db}) (i \pa_{\a \da} \cwb_{\db}) 
                  \dua \f (\bar{D} \fb)^2 \over \f^3 \fb^4}     \ret 
& & \qquad - { (\dua \dbua \fb) (i \pa_{\b \db} \dla \f) 
                   \cwb^{\db} \dbla \fb \cw^{\b} \over \f^3 \fb^4} 
 - {3 \over 2} { (\dua \dbua \fb) (i \pa_{\b \da} \dla \f) 
                    D^{\b} \f (\cw)^2 \over \f^4 \fb^3}         \ret
& & \qquad - { (\dua \dbua \fb) (i \pa_{\a \db} \cwb_{\da})
                   \cwb^{\db} (\bar{D} \fb)^2 \over \f^2 \fb^5} 
 + {1 \over 2} { (\dua \dbua \fb) (i \pa_{\a \da} \cwb_{\db}) 
                  \bar{D}^{\db} \fb \cw^{\b} D_{\b} \f \over \f^3 \fb^4} \ret
& & \qquad - { (\dua \dbua \fb) (i \pa_{\a \db} \cw_{\b}) 
                \bar{D}^{\db} \fb \cwb_{\da} D^{\b} \f \over \f^3 \fb^4}
 - { (\dbua \cwb^{\db}) (i \pa_{\a \dg} \dbla \fb) 
         \dua \f \bar{D}_{\db} \fb \cwb^{\dg} \over \f^3 \fb^4}  \ret
& & \qquad \qquad \qquad \qquad \qquad + \;\; c.c. \nn
\eeqs    
              
 We can evaluate now the $N=1$ components of $S^{(6)}_{eff} (1/\t)$ in 
(\ref{6act}). Dropping again any term depending on auxiliary superfields 
we obtain

\beqs
 S^{(6)}_{eff} (1 / \t) & = & {g^2 c^2 \over (8 \pi)^2} \int d^4 x \, d^4 \th 
  \;\; 3 { \shalf (\bar{D} \cwb)^2 \; \shalf (D\bar{D} \fb)^2 
               \over \f^2 \fb^4} 
 + { (\dbua \cwb_{\db}) (\dua \bar{D}^{\db} \fb)
  ( i \pa_{\a \dg} \dbla \fb) \cwb^{\dg} \over \f^2 \fb^4}     \ret
& & \qquad \qquad + 6 \; {\shalf (\ddbfb)^2 (\dbua \cwb^{\db}) 
                   \cwb_{\db} \dbla \fb \over \f^2 \fb^5}
 + 4 {\shalf (\bar{D} \cwb)^2 (\dua \dbua \fb) \cwb_{\da} \cw_\a 
                     \over  \f^3 \fb^4}                        \ret
& & \qquad \qquad - 20 \; { \shalf (\ddbfb)^2 (\cwb)^2 (\bar{D} \fb)^2 
                     \over \f^2 \fb^6 }
 + 8 { (\dbua \cwb^{\db}) (\dua \bar{D}_{\db} \fb) 
             \cw_{\a} \dbla \fb (\cwb)^2 \over \f^3 \fb^5 }    \ret
& & \qquad \qquad + \; {1 \over 2} { (\pa \bar{D} \fb)^2 (\cwb)^2 \over 
       \f^2 \fb^4} 
    - 2 { (\dua \dbua \fb) (i \pa_{\a \db} \bar{D}_{\da} \fb) (\cwb)^2
          \bar{D}^{\db} \fb \over \f^2 \fb^5}  \ret
& & \qquad \qquad - \; 6 \; { \shalf (\bar{D} \cwb)^2 (\cwb)^2 (\cw)^2 
                              \over \f^4 \fb^4 } \ret
& & \qquad \qquad \qquad \qquad + \;\; c.c. 
\label{6act_n1}
\eeqs

 We can see that all the terms in this action except the second 
and the last three have precisely the correct numerical coefficients for 
them to be absorbed in the redefinition of the superfields appearing in 
$S^{(2)}_{eff} + S^{(4)}_{eff}$. Using part of the last term in the 
redefinition all we are left with is

\beqs
 S^{(6)}_{eff} - O^{(6)} - \ep S^{(6)}_{fermi} & = & 
 {g^2 c^2 \over (8 \pi)^2} \int d^4 x \, d^4 \th \: 
   - 2 \; { \shalf (\bar{D} \cwb)^2 (\cwb)^2 (\cw)^2 \over \f^4 \fb^4 } 
 + {1 \over 4} { (\pa \bar{D} \fb)^2 (\cwb)^2 \over \f^2 \fb^4} \ret
 & & \qquad \qquad + \; {1 \over 2} { (\dbua \cwb_{\db}) 
     (\dua \bar{D}^{\db} \fb) ( i \pa_{\a \dg} \dbla \fb) \cwb^{\dg} 
       \over \f^2 \fb^4} \ret
 & & \qquad \qquad - \; { (\dua \dbua \fb) (i \pa_{\a \db} \bar{D}_{\da} \fb) 
           (\cwb)^2 \bar{D}^{\db} \fb \over \f^2 \fb^5} \ret
& & \qquad \qquad \; + \; c.c. 
\eeqs

The first term is precisely the six derivative contribution we are 
looking for in the redefined $N=1$ action. The last three we expect to 
cancel against similar terms in $S^{(6)}_{eff} (1/ |\t|)$. In any case 
such contributions contain products of velocities and gauge field 
strengths, and therefore do not affect our comparison with the gauge 
part of the DBI action.  

 Since we do not know the exact form $S^{(6)}_{eff} (1/ |\t|)$ we can 
only try to guess the operators present in this piece of the effective 
action and fix the coefficients by matching their $N=1$ components with 
$O^{(6)} (1/ |\t|)$. This is a difficult task, further complicated by 
the ambiguity of integration by parts. It is worth mentioning that the 
$N=1$ components of the operators in (\ref{simp_ops}) are different 
from those in (\ref{6act_n1}) and therefore our partial result is 
unchanged by the $\cp^{ab}$ terms. The analysis of these 
contributions will be presented in a future publication.


\begin{thebibliography}{9}

\bibitem{MO}{C. Montonen, D. Olive, \PLB{72} (1977) 117.}

\bibitem{S}{ A. Sen, \PLB{329} (1994) 217, hep-th/9402032.}

\bibitem{sw}{ N.~Seiberg and E.~Witten, \NPB{426} (1994) 19, hep-th/9407087; 
 Erratum-ibid \NPB{430} (1994) 485; \NPB{431} (1994) 484, hep-th/9408099.}

\bibitem{book}{ S.J.~Gates, M.T.~Grisaru, M.~Ro\v{c}ek and W.~Siegel 
 {\em Superspace} (Benjamin-Cummings 1983).}

\bibitem{ni}{ B.~de Wit, M.T.~Grisaru and M.~Ro\v{c}ek, 
 \PLB{374} (1996) 297, hep-th/9601115.}

\bibitem{DS}{M. Dine and N. Seiberg, \NPB{458} (1996) 445, 
 hep-th/9707057.}

\bibitem{rikard_vipul}{V. Periwal and R. von Unge, \PLB{430} (1998)
71, hep-th/9801121.}

\bibitem{GR}{F. Gonzalez-Rey and M. Ro\v{c}ek, \PLB{434} (1998) 303,
 hep-th/9804010.}

\bibitem{kuz}{I.L. Buchbinder and S.M. Kuzenko, Mod. Phys. Lett. {\bf A 13} 
(1998) 1623; hep-th/9804168.}

\bibitem{dorey}{N. Dorey, V.V. Khoze, M.P. Mattis, J. Slater and W.A. Weir, 
 \PLB{408} (1997) 213, hep-th/9706007.} \\
{D. Bellisai, F. Fucito, M. Matone and G. Travaglini, \PRD{56} (1997) 5218, 
 hep-th/9706099.}

\bibitem{henning}{ M. Henningson, \NPB{458} (1996) 445, hep-th/9507135.}

\bibitem{menzi}{L. Mezincescu, JINR preprint P2-12572 (1979).\\
S. J. Gates and W. Siegel, \NPB{189} (1981) 295.\\
P. S. Howe, K. S. Stelle and P. K. Townsend, \NPB{236} (1984) 125.}

\bibitem{witten}{E. Witten, \NPB{460} (1996) 335, hep-th/9510135.}

\bibitem{gib_rash}{G.W. Gibbons, D.A. Rasheed, \NPB{454} (1995) 185,
hep-th/9506035.}

\bibitem{maldac}{J. M. Maldacena, Nuc. Phys. Proc. Suppl. {\bf 68} (1998) 17,
hep-th/9709099.}

\bibitem{dkps}{M. R. Douglas, D. Kabat, P. Pouliot and S. H. Shenker,
\NPB{485} (1997) 85, hep-th/9608024.}

\bibitem{aga}{M. Aganagic, J. Park, C. Popescu and J. Schwarz, \NPB{496} 
(1997) 215, hep-th/9702133.}

\bibitem{park_rey}{J. Park and S.-J. Rey, hep-th/9810154.}

\bibitem{chep}{I. Chepelev and A.A. Tseytlin, \NPB{511} (1998) 629,
hep-th/9704127.}

\bibitem{hor_stro}{G. Horowitz and A. Strominger, \NPB{360} (1991) 197.\\
 M.J. Duff and J.X. Lu, \PLB{273} (1991) 409.}

\bibitem{iuri}{I. Chepelev and A.A. Tseytlin, \NPB{515} (1998) 73, 
hep-th/9709087.}

\bibitem{esko}{E. Keski-Vakkuri and P. Kraus, \NPB{518} (1998) 212,
hep-th/9709122.}

\bibitem{paban}{S. Paban, S. Sethi and M. Stern, JHEP {\bf 06} (1998)
012, hep-th/9806028; hep-th/9805018.} 

\bibitem{andreev}{O.D. Andreev and A.A. Tseytlin, \NPB{311} (1988) 205.}

\bibitem{tseyt}{A.A. Tseytlin, \NPB{469} (1996) 51, hep-th/9602064.}

\bibitem{mald_conj}{J. Maldacena, Adv. Theor. Math. Phys {\bf 2} (1998) 231,
hep-th/9711200.}

\bibitem{douglas_taylor}{M. Douglas and W. Taylor, hep-th/9807225.}

\bibitem{trivedi}{S. R. Das and S. P. Trivedi, hep-th/9804149.}

\bibitem{diego}{D. Bellisai, Research in progress.}

\bibitem{bagr}{T. Banks and M. Green, JHEP {\bf 05} (1998) 002, 
hep-th/9804170.}

\bibitem{kuz2}{E.I. Buchbinder, I.L. Buchbinder, S.M. Kuzenko, hep-th/9810239.}

\bibitem{rik_lowe}{D.A. Lowe, R. von Unge, JHEP {\bf 11} (1998) 014, 
hep-th/9811017.}

\bibitem{rocek_tseyt}{M. Ro\v{c}ek and A. Tseytlin, hep-th/9811232.}

\bibitem{nonlin}{F. Gonzalez-Rey, I.Y. Park and M. Ro\v{c}ek, hep-th/9811130.}




\end{thebibliography}
\end{document}